# Phase state dependent current fluctuations in pure lipid membranes

B. Wunderlich[1], C. Leirer[1,3], A-L. Idzko[1], U. F. Keyser[2], A. Wixforth[1], V.M. Myles[1], T. Heimburg[3], M.F. Schneider[1*]

[1] University of Augsburg, Experimental Physics I, D-86159 Augsburg, Germany
[2] Universität Leipzig, Experimentalphysik I, D-04103 Leipzig, Germany
[3] Niels Bohr Institute, University of Copenhagen, DK - 2100 Copenhagen Ø, Denmark

**Running Titel: Membrane Current Fluctuations**

29.07.2008

*\* Corresponding author:*

*Matthias F. Schneider*

*Phone: +49-821-5983311, Fax: +49-821-5983227*

matthias.schneider@physik.uni-augsburg.de

*University of Augsburg, Experimental Physics I,*

*Biological Physics Group*

*Universitaetstr. 1*

*D-86159 Augsburg, Germany*

## Abstract

Current fluctuations in pure lipid membranes have been shown to occur under the influence of transmembrane electric fields (electroporation) as well as a result from structural rearrangements of the lipid bilayer during phase transition (soft perforation).

We demonstrate that the ion permeability during lipid phase transition exhibits the same qualitative temperature dependence as the macroscopic heat capacity of a D15PC/DOPC vesicle suspension. Microscopic current fluctuations show distinct characteristics for each individual phase state. While current fluctuations in the fluid phase show spike-like behaviour of short time scales (~ 2ms) with a narrow amplitude distribution, the current fluctuations during lipid phase transition appear in distinct steps with time scales in the order of ~ 20ms.

We propose a theoretical explanation for the origin of time scales and permeability based on a linear relationship between lipid membrane susceptibilities and relaxation times in the vicinity of the phase transition.

**Introduction**

The permeability of cell membranes to ions, proteins and other transmitters and the regulation of these processes are of crucial importance in maintaining basic cell functions. In this context ion channel proteins have been found to play a major role in the regulation of ionic transport across the membrane (1). Besides the characteristics of the channel proteins in itself an influence of the membrane phase state on channel function has been identified (2,3), indicating the importance of the lipid bilayer for transmembrane currents. Boheim et al., for example, reported on the slowing down of protein induced ion channels during the lipid phase transition (32,33). The first indication of thermodynamically induced changes in permeability was given by Papahdjopoulus et al. already in 1973 (4), who found a maximum leakage of radioactive labelled sodium ions during lipid phase transition.

The application of high electric fields (typically ~ $10^7$ to $10^8$ V/m) presents another mechanism for overcoming the lipid membrane barrier, inducing current fluctuations in the absence of channel forming peptides (electroporation) (5-9).

A third mechanism (in addition to ion channels, pores and electroporation) to cross the lipid membrane barrier has been elucidated by Antonov et al. (1980) (10). They reported that quantized ion-conducting channels appear in unmodified 1,2-Distearoyl-*sn*-Glycero-3-Phosphocholine (DSPC) membranes close to the phase transition and demonstrated more recently, that the conductance of different ions is in accordance with the well known Hofmeister series (11). Similarly, Kaufmann and Silmann (12) as well as (2,13-16) were able to demonstrate a close relationship between the physical state of the bilayer and ion current fluctuations. This relationship is also the subject of an accompanying paper by Blicher et al. (31).

In this paper we study the origin of ion current fluctuations in protein free lipid bilayers by a thorough comparison of timescales and amplitudes of ion current fluctuations with the heat capacity profile of lipid membranes. While current fluctuations in the fluid phase appear as short (~2ms) spikes, they last for ~20ms as discrete step-like currents in the phase transition regime including long lasting events of ~100ms. Based on a model recently proposed by Grabitz et al. (17), we suggest that the increased time scales at $T_m$ have their origin in the extended relaxation times generally found during lipid membrane phase transitions and originating from the flat thermodynamic potential and consequent weak restoring forces driving the system back to its equilibrium near $T_m$.

**Materials and Methods**

Lipids 1,2-Dioleoyl-*sn*-Glycero-3-Phosphocholine (DOPC, $T_m$ = *-4°C*) and 1,2 Dipentadecanoyl-*sn*-Glycero-3-Phosphocholine (D15PC, $T_m$ = *33°C*) dissolved in Chloroform were purchased from Avanti Polar Lipids (Birmingham, Al. USA) and used without further purification. Hexadecane was dissolved in Pentane to achieve a final concentration of 2,5% for the prepainting solution.

We performed current measurements using a patch clamp amplifier (Cornerstone Series) from Dagan Corp. (Minneapolos, Min., USA). The temperature was controlled during the experiments with the aid of a standard heat bath (Julabo GmbH, Seelbach, Germany). Heat capacity profiles of small unilamellar vesicles (SUVs) were recorded using a VP-DSC calorimeter from Microcal Inc. (Avestin Inc., Ottawa, Canada).

To obtain lipid mixtures, lipids dissolved in chloroform were mixed in the desired proportion. Aqueous solution of lipids for calorimetric measurements were obtained as described previously (18). The experiments on the artificial membranes (BLM) were performed by using a setup consisting of two Teflon chambers separated by a Teflon foil of thickness 25μm with a hole of diameter 150μm in it. The hole was produced by drilling it into the foil. Lipid membranes were obtained by prepainting the Teflon foil with the Hexadecan solution and 15min of waiting until the Pentane was evaporated. Subsequently, both chambers were filled up to a level sufficiently far below the hole with a 200 mM NaCl solution. 10 μl of lipid solution was spread on the water surface and after waiting time of 15 min the electrolyte level was raised above the hole. Prior to experiments, the bilayer was characterized by measuring the resistance and the capacity of the membrane.

**Results and Discussion**

**Theory**

In thermodynamics, the different functions of state (enthalpy, energy etc), the susceptibilities (heat capacity, compressibilities etc) and the response times are coupled via the fluctuation-dissipation theorem (19). For lipid membranes, it has been shown by Grabitz et al. (17) and Seeger et al. (20) that the relaxation time scale $\tau$ after a pressure perturbation is proportional to the excess heat capacity $\Delta c_p$ such that

$$\tau = \frac{T^2}{L}\Delta c_p \qquad (1)$$

where T is the temperature in Kelvin and L is a phenomenological coefficient independent of temperature (see Grabitz et al., 2002 (17), and Seeger et al., 2007 (20), for details). For DMPC MLV for instance, L=6.6·10$^8$ J K/mols and for DMPC LUV it is L=15.7·10$^8$ J K/mols. This equation implies that relaxation is slow if the system is at the heat capacity maximum, which should be very pronounced in lipid membranes exhibiting a phase transition. For multilamellar DPPC vesicles, this time can be as long as 30 seconds. According to Onsager (21,22) each fluctuation can be considered as a perturbation of the system away from the entropy maximum. Therefore, fluctuation lifetimes and relaxation times are the same. According to Einstein (23), fluctuations in the thermodynamic variable $n$ are related to the curvature of the thermodynamic potential $G$ by

$$\langle \partial n^2 \rangle = -k_B T \left( \frac{\partial^2 G}{dn^2} \right)^{-1},$$

where $k_b$ is the Boltzmann constant and $T$ the temperature.

This can be used to show that heat capacity and fluctuations in enthalpy $H$ are proportional,

$$\langle \partial H^2 \rangle = \langle H^2 \rangle - \langle H \rangle^2 = RT^2 c_p,$$

around the entropy maximum ($R$ is the gas constant). Along the same lines one can show that the lateral compressibility $\kappa_T$ is given by

$$\langle \partial A^2 \rangle = ART\kappa_T, \tag{2}$$

i.e., it is proportional to the fluctuations in area $A$. Heimburg (24) and Ebel et al. (25) have shown that changes in heat, in volume and in area in lipid transitions are proportional functions. This leads to the relation

$$\Delta \kappa_T^A = \frac{\gamma^2 T}{A} \Delta c_p \tag{3}$$

meaning that the lateral compressibility changes are proportional to the excess heat capacity ($\gamma$ is a material constant of the order of 1 m²/J (24)). This further implies that it does not matter whether one considers a perturbation in enthalpy (after a temperature jump) or a perturbation in volume or area. The fluctuations in volume and area have the same lifetimes as those of the heat.

Pore formation in absence of channel forming proteins occurs due to thermal area fluctuations. According to Nagle and Scott (26), the work to create a pore in the membrane is proportional to its lateral compressibility, which is proportional to the excess heat capacity (Eq. 3). As a consequence, pore formation is facilitated in the melting regime. Membranes are expected to be more permeable at the heat capacity maximum (for details see ref. 31 in this issue). Equating the fluctuation lifetime with the lifetime of pores, this implies that the pore opening times are proportional to the heat capacity. This relation has already been discussed in Seeger et al. (20), and in this paper we will demonstrate that this relation leads to reasonable predictions. Both permeability and mean pore open time are at maximum in the chain melting transition. Experimental values for pore lifetimes are close to those estimated from Eq. (1).

**Permeability changes during lipid membrane phase transition**

In order to study the temperature dependence of lipid bilayers, a BLM was prepared at 40°C well above the phase transition of the lipid mixture used (5/95 DOPC/D15PC). Prior to the BLM experiments the heat capacity of the lipid mixture has been measured (inset of Fig. 1) revealing a melting transition at $T_m \approx 30°C$. To assure bilayer formation the electric capacitance of the membrane film has been monitored. The bilayer was then left to equilibrate for 15 min. After film formation, the stability of the bilayer was tested by a stepwise increase of the clamp voltage until significant current fluctuation occurred (threshold voltage). Subsequently, the applied voltage $V_M$ was decreased until the current fluctuations disappeared, leaving the system close to the conducting state. After 10 min waiting for equilibration, the temperature was lowered at a rate of 0,2 °C/min at constant clamp voltage. Fig. 1 shows the overall trans membrane conductivity (averaged over 60 seconds) at $V_M = 400$ mV as a function of temperature. At T=36°C the trans membrane conductivity begins to rapidly rise to reach a maximum at T≈ 31,5°C. Clearly, ion permeability and heat capacity follow the same qualitative temperature dependence with a full width at half height (FWHH) of 3-6 K. The slight shift of $I_{tot}$ (T) towards higher temperatures is probably due to the different sample preparations. After sonification SUVs are believed to be under some tension while BLMs are rather relaxed. Tension, however, tends to increase the membrane area thereby supporting a lowering in $T_m$. Finally, uncontrollable lipid accumulation around the septum might be an additional source of changes in $T_m$. We were able to observe the same correlation between heat capacity and ion permeability for D15PC/DMPC mixtures (see supplementary material). However, the introduction of DOPC in our lipid mixtures significantly increased the stability of the planar lipid membrane. This can also be seen in fig. 2b, where the bilayer resists membrane potentials as high as 1000mV.

One possible origin of the increase in overall permeability is the increasing number of lattice defects at the gel/fluid phase boundary. This was suggested by several theoretical findings (27) and is supported by the work of Papahadjopoulos et al. (4), who found a maximum leakage of radioactive labelled sodium ions during lipid phase transition. An alternative approach to explain the increased permeability in the phase transition region was given by Kaufmann (28) as well as by Nagel and Scott (26) and Blicher et al. (31) who address the changes in lateral compressibility $\kappa_T$ and their relation to the fluctuations in area $\langle \partial A^2 \rangle$ (Eq. 2). Considering such area fluctuations as a source of defects excellently agrees with the correlation between heat capacity and conductivity.

**Current characteristics strongly depend on lipid phase state**

To unravel the origin of the transient conductivity behaviour, we prepared BLM's from a DOPC-D15PC (5:95) mixture both in the fluid phase (40°C) and in the phase transition region (27°C). After a stepwise increase in $V_M$ until values of which current fluctuations were observed, the voltage $V_M$ was again slightly reduced resulting in stable current fluctuations over several minutes.

In the fluid phase (Fig. 2a, *T = 33°C, $V_M$ = 100 mV*), spike-like events were observed and did not change its characteristics at even higher temperatures. The current fluctuations in the phase transition regime (Fig. 2b, *T = 31,5°C, $V_M$ = 1000 mV*) however, appear rather step-like and quantized with longer opening times. Again, the bilayer exhibited very high stability and the relation between applied voltage and initial resistance will be addresses in a separate issue. For the DMPC/D15PC mixture for example, strong current fluctuations were found around 500mV and below (see supplementary). An analysis of the current amplitudes results in a rather broad distribution with two small maxima around 12 pS and 24 pS in the fluid state (Fig 3a). In the phase transition regime these maxima become very pronounced (Fig. 3b). Also the gel state exhibited some current fluctuations, which, however either slowly disappeared when waiting for extended times or led to membrane rupture (see supplementary data). Distinct quantized current fluctuations in lipid membranes have been reported before (2,10-16,26,31) ranging between ~1 pA and ~1 nA close to the phase transition temperature. However, no clear relation between current amplitudes, distribution, and fluctuations for the different thermodynamic states of the lipid membrane has been given, so far.

The existence of current fluctuations in both states suggests that the nature of the underlying mechanism for these current fluctuations is the same, but the specific characteristics seem to be determined by the physical properties of the membrane. In agreement with our results, we believe, that fluctuations in area $\langle \partial A^2 \rangle$ are the origin of lattice defect formation and hence permeating ions. However, the origin of the step-like shape of the current fluctuations remains unclear yet is probably related to the physical properties of the liquid crystal. In principle, and based on our experimental results, one could model the pore-size necessary for an ion to pass the membrane. For a one-step fluctuation (Fig. 2b) this suggests a single pore diameter of ~ 1 nm, using a cylindrical pore approximation assuming free instead of surface bound water within the pore.

**Time scales of current fluctuation increase during phase transition**

To unravel the relationship between thermodynamic behaviour and the observed fluctuations, we analyzed the typical timescales of the opening times. Peaks were identified and characterized with a threshold criterion on the same set of data used to describe the conductance characteristics above (Fig. 3). In Figure 4, we show the resulting data for the fluid phase (circles) and the phase transition (triangles). The logarithmic plot already reveals that the opening times are significantly increased within the phase transition regime. In the fluid phase we found a mean average opening time in the order of 3 ms (exponential fit). The behaviour alters significantly when the phase transition region is entered. The mean average timescales shift by almost an order of magnitude to ~20 ms including long lasting states of up to ~100 ms. We also observed a broad variety of events including short events in the transition region and long fluctuations in the fluid state during measurements over extended periods of times. This is also reproduced by the broad distributions shown in Fig. 3 and reflects uncontrollable variations in the lipid membrane structure and dynamics like for example the pool of lipids accumulating around the septum or the degree of asymmetry. However, the tendency towards increased time scales during lipid phase transitions is obvious and reproducible in every experiment. Importantly, these enhanced timescales are in line with the increasing relaxation times of the membrane in this regime (17) and originate most likely from the small restoring forces of the flat thermodynamic potential around $T_m$. Experimentally, increased relaxation times during the main phase transition of artificial membranes have been reported before by Tsong et al. (29) as well as by Blume et al. (30). Since the relaxation times reflect the reorganization of lipids in the membrane, one would expect that pore nucleation, pore opening, and closing respectively, follow the same time evolution as lipid relaxation.

Taking the phenomenological coefficient $L$ for DMPC MLV (as given above) and numbers for $c_p$ from Fig. 1 (at ~31°C), Eq. 1 yields lifetimes of 165ms, while using the L from DMPC LUV yields lifetimes of 73 ms. Considering the transition halfwidth of only 2°C, this is very close to the experimentally found average lifetime of 20 ms (Fig. 4).

**Conclusion**

The behaviour of current fluctuations in an unmodified phospholipid membrane under a constant voltage was found to be strongly sensitive to the phase state. Whereas fluctuations in the fluid phase reveal an unordered pattern on short timescales, in the phase transition region "quantized" fluctuations on much longer timescales occur. Using the fluctuation dissipation theorem and the linear relationship between heat capacity and area compressibility we were able to predict the correct timescales and identify weak restoring forces of the flat thermodynamic potential as the origin of the extended time scales during lipid phase transition. Therefore, this paper demonstrates the thermodynamic potential of lipid membranes to be the physical origin of lipid membrane current fluctuations.

**Acknowledgments**

We would like to thank Dr. K. Kaufmann (Göttingen) for very helpful discussions and would like to draw the readers attention to his earlier work [s Ref. 28]. MFS likes to personally thank K. Kaufmann who inspired him to work in this field and outlined the fundamental problems of transport across lipid membranes and its thermodynamic origin.

Financial support by the Deutsche Forschungsgemeinschaft DFG (SPP 1313) and the German Excellence Initiative via the "Nanosystems Initiative Munich (NIM)" is gratefully acknowledged. CL likes to thank the Bavarian Science Foundation for financial support.

**Figure Captions**

Figure 1: Relative conductance of a D15PC-DOPC (95/5) mixture as a function of temperature at $V_M$ = 400mV. A maximum is observed between 35°C and 25°C. The current trace clearly correlates with the heat capacity profile (see inset). The maximum corresponds to a conductivity of  $2 \cdot 10^{-7}$S/cm². Conductivity measurements outside the transition region in the fluid (squares) and gel phase (triangles) have been performed on individual BLMs and are further described in the supplementary data.

Fig 2: Typical current traces of a D15PC-DOPC (95/5) (a) in the fluid phase (*33°C, $V_M$ = 100~mV*) and  the phase transition regime (*31.5C, $V_M$ = 1000 mV*) (b). In the fluid phase spike-like current fluctuations on short times scales are observed. In the phase transition regime however, quantized fluctuations appear at longer time scales. Note that the typical time scales are strongly increased as compared to a).

Fig 3: Conductivity histogram of current fluctuations in the fluid phase ($T = 33°C$, $V_M = 100$ mV) (a) and the phase transition regime ($T = 31.5°C$, $V_M = 1000$ mV) (b). The peak at $G = 0$ pS corresponds to the baseline. In the fluid phase no significant peaks appear. In the phase transition (b) regime distinct maxima around $G = 12$ pS and $G = 24$ pS appear, with some additional substructure around $G = 20$ and $G = 27$ pS. Data were collected from three individual bilayers by analyzing a 200 s long trace. The observed maxima agree in all measurements within a range of 20%.

Fig 4: Logarithmic plot of the timescales of the current fluctuation in the fluid phase (triangles) and the phase transition regime (circles). The transmembrane currents have been measured under constant membrane potential $V_m$. The timescales of the fluid phase follow an exponential decay while those in the transition can only be approximated by a double exponential decay. The average lifetimes $\tau$ centre around $\tau = 3$ms in the fluid and $\tau = 20$ms in the phase transition. This is in good agreement with our theoretical prediction.

**Figures**

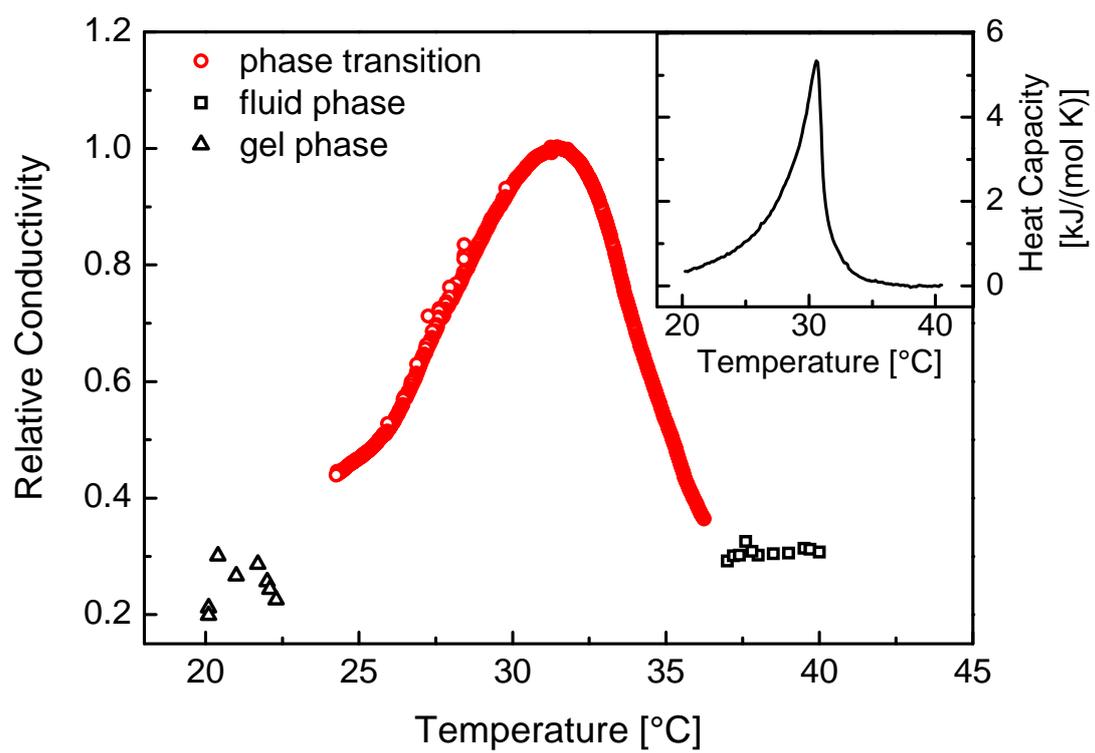

**Figure 1**

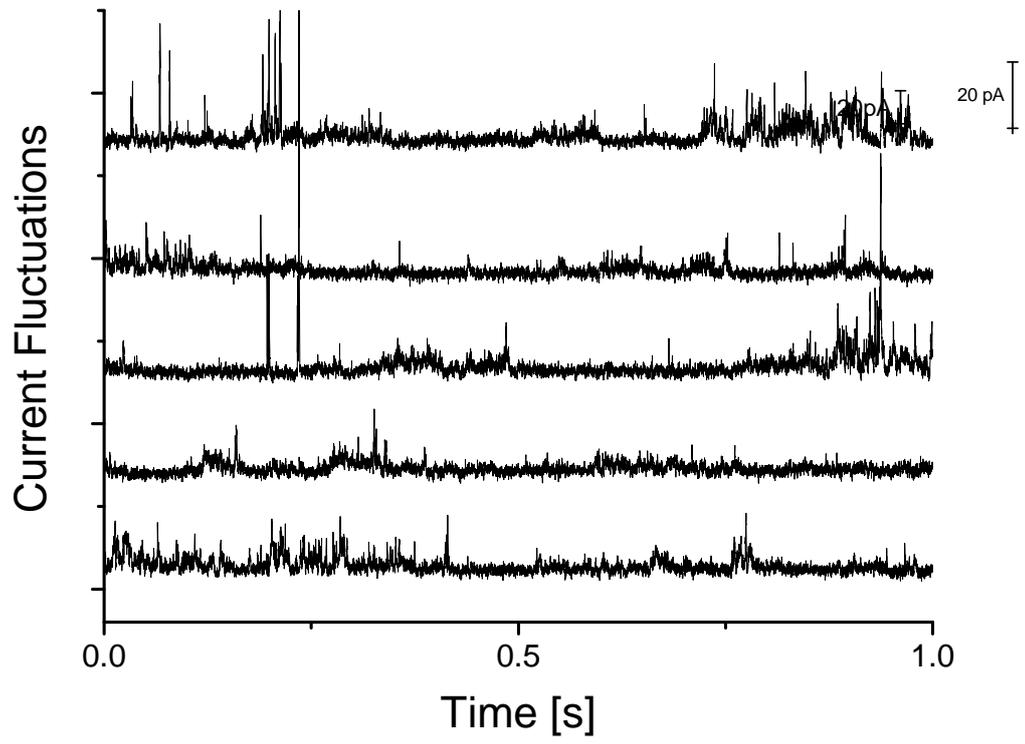

**Figure 2a**

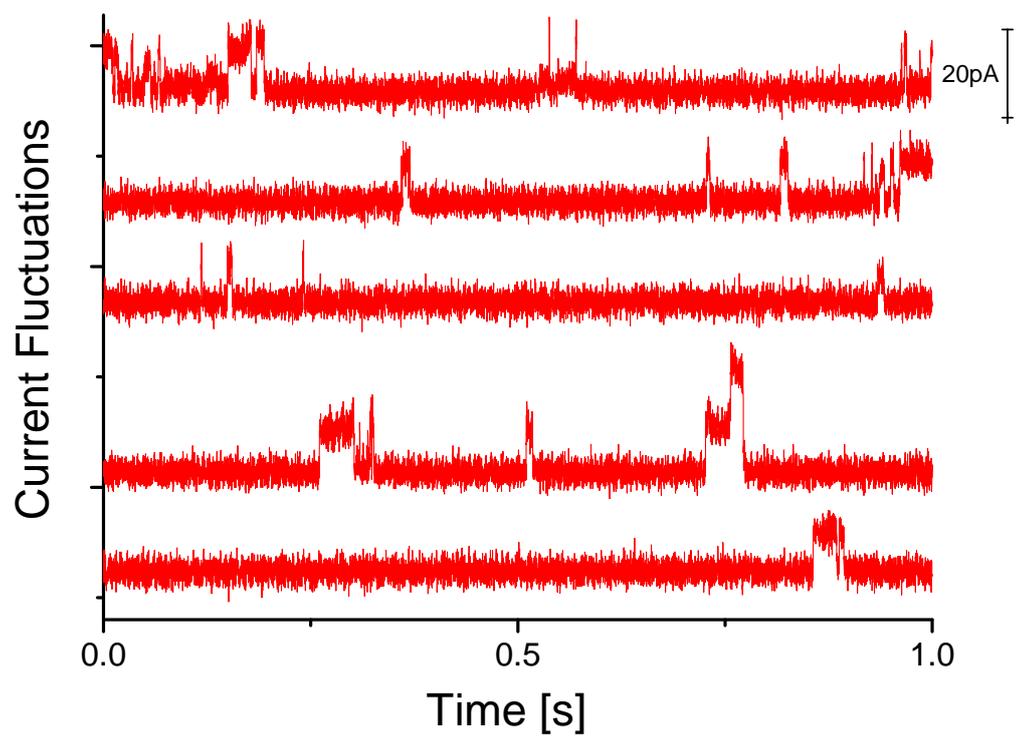

**Figure 2b**

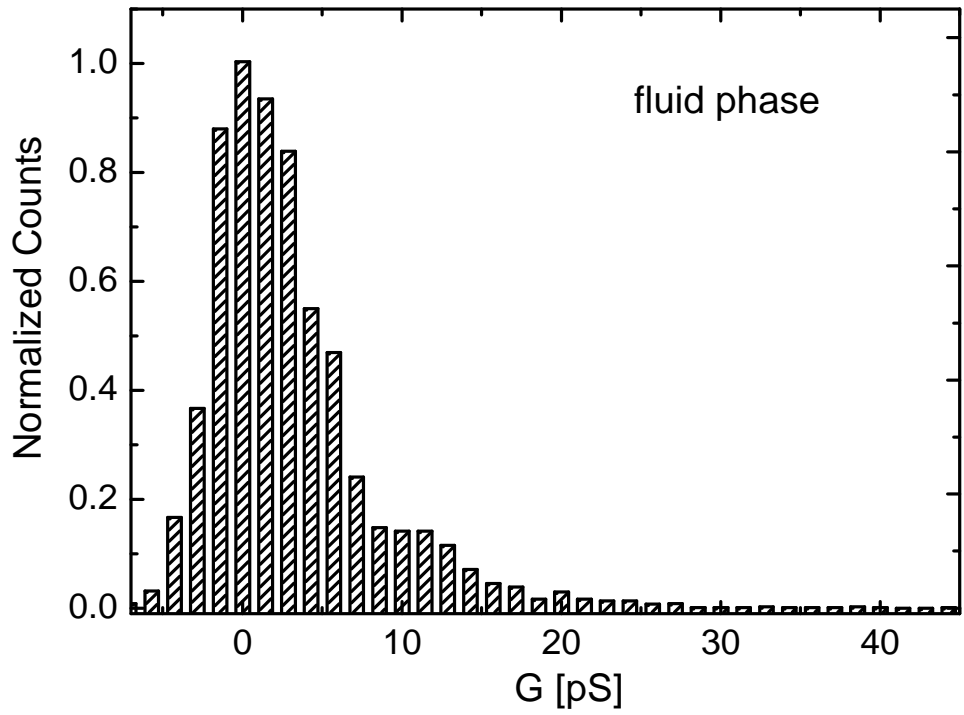

**Figure 3a**

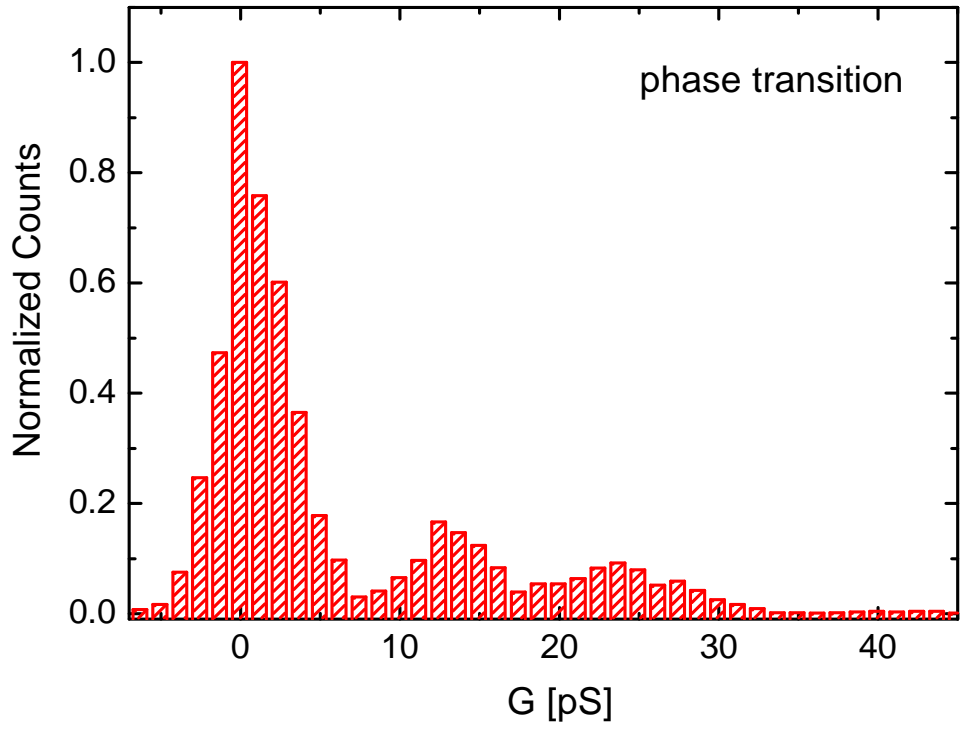

**Figure 3b**

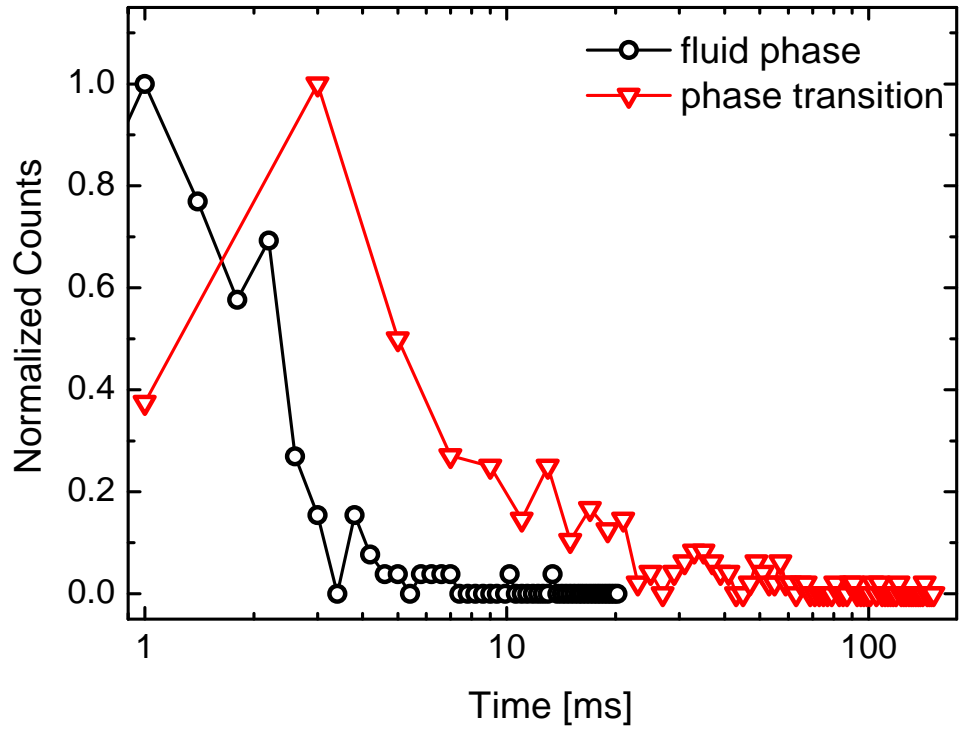

**Figure 4**